## Article title

MEMSDuino: An Arduino-Based MEMS Switch Controller

## Authors
Lafe Spietz*[1], Adam Sirois[1], Nathan Flowers-Jacobs[1], Steve Waltman[2], Samuel Benz[1], and Peter Hopkins[1]

## Affiliations

[1]National Institute of Standards and Technology, 325 Broadway, Boulder, CO, 80305, USA [2]Google Quantum AI, Santa Barbara, CA, USA

## Corresponding author's email address

Lafe.f.spietz@nist.gov

## Abstract

Radio frequency cryogenic switches are a critical enabling technology for quantum information science for both calibration and high throughput testing of samples.  Traditionally, solenoid-based switches have been used [1,2], but a transition is being made to MEMS-based (Micro Electro Mechanical Systems) switches due to their lower power dissipation and smaller size, and to minimize the risk that solenoid switches tend to produce current pulses that destroy expensive cryogenic amplifiers and can cause electrostatic damage to devices.  These MEMS switches require a 90-volt signal to be applied to the control lines to determine the state of the switches.  Switches exist that have built-in CMOS-based (Complimentary Metal Oxide Semiconductor) control electronics to drive the 90 V, but these do not work at the cryogenic temperatures used in quantum information science.

There is no currently available room temperature control system with direct control of the switches.  The instrument presented here is a 19-inch rack-mount controller for a cryogenic MEMS switch network that allows a human operator to see the state of the switch via a row of clearly marked indicator lights and to change the state manually via buttons on an LED-based indicator board or automatically via Python-based serial port commands to the Arduino, an open source microcontroller platform available from multiple vendors. The design can also be modified to control other switches that require either a large voltage or current to switch.

## Specifications table

| | |
|---|---|
| Hardware name | *MEMSDuino* |
| Subject area | • Physics<br>• Educational tools and open-source alternatives to existing infrastructure<br>• General |
| Hardware type | • Measuring physical properties and in-lab sensors<br>• Electrical engineering and computer science |
| Closest commercial analog | No commercial analog is available |
| Open source license | Public Domain |
| Cost of hardware | ~$60 - $120 |
| Source file repository | *If you've uploaded your source files to an approved repository (OSF, Mendeley Data or Zenodo)* write the *DOI URL here.*<br>*For example:* http://doi.org/10.17605/OSF.IO/WGK7Q |
| OSHWA certification UID *(OPTIONAL)* | *We encourage you to apply for a free OSHWA Certification, which confirms your work is open-source compliant.*<br>*If certification has been acquired, insert the OSHWA UID here. For example: "CH000005". In your OSHWA certification project description, include a link to your HardwareX publication and the tag #HX.*<br>*If you haven't acquired certification, please delete this row of the specifications table.* |

## 1. Hardware in context

Measurements and calibrations using low-power (below -60 dBm) radio frequency (RF) signals are of increasing interest due to the growth of quantum information technology, dark matter studies, and fundamental microwave quantum optics studies [3-6]. To observe RF signals in the quantum regime, the circuits being measured must be cooled to temperatures well below $hf/k_B$, where h is the Planck constant, f is the frequency, and $k_B$ is the Boltzmann constant. For the microwave frequency range (1- 10) GHz typical in superconducting quantum information technology, if we want $k_BT$ to be more than 10 times below hf, this means we need temperatures in the range from (5-50) mK. This is generally accomplished with a dilution refrigerator (see Figure 1).

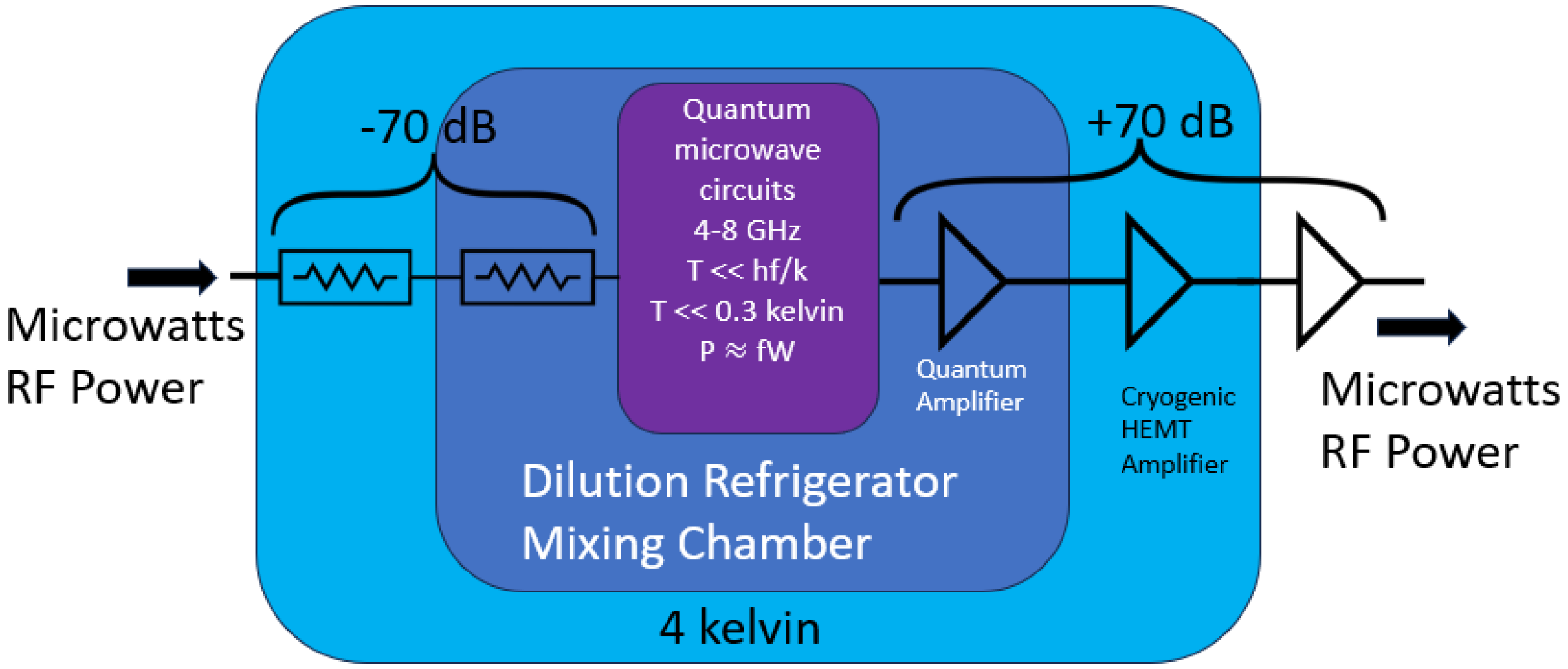

**Figure 1**. Schematic of typical quantum information experiment. Typically, there are about 7 orders of magnitude of attenuation on one side and about the same amount of gain on the other side in order to achieve appropriate signal-to-noise to detect the very small quantum signals.

If a circuit is to thermalize to these low temperatures, power going into the dilution refrigerator needs to be attenuated by several orders of magnitude. To not overwhelm the extremely small quantum signals (<1 femtowatt) with noise, these circuits generally also have cryogenic quantum-noise-limited parametric and HEMT (High Electron Mobility Transistor) amplifiers inside the dilution refrigerator, which amplify the signals on the way out of the cryostat by several orders of magnitude in power and add to the heat load on the system.

If researchers want to measure large numbers of samples in a single signal path, calibrate the noise, or calibrate the scattering parameters, they need to have the ability to switch a single signal line between a choice of multiple lines and devices under test while the apparatus is cold. At ambient temperatures, there are many solutions for this, but at low temperatures, the three known solutions are 1) solenoid-based cryogenic electro-mechanical switches, 2) piezo-based probe manipulators, and 3) MEMS-based electro-mechanical switches. The solenoid-based switches require large current pulses that dissipate heat and warm up the cryostat, requiring about an hour after each switching event for the system to cool back down to base temperature. These switches can also damage the cryogenic HEMT amplifiers, costing several thousand dollars to replace with potentially long lead times, as well as destroying Josephson Junction devices and other sensitive electronics. The piezo-based probe manipulators are the first choice for performing cryogenic on-wafer calibrations but also cause heating, are expensive, slow, and cumbersome to use, and their large physical footprint makes them incompatible with many of the EMI and magnetic shield enclosures required in quantum information science experiments.

This work presents a flexible, inexpensive, and simple control system for MEMS-based microwave switches from Menlo Micro (part number MM5130-03NDB-TR). [FOOTNOTE: Certain commercial instruments are identified in this article to facilitate understanding. Such identification does not imply recommendation or endorsement by NIST, nor does it imply that the materials or equipment that are identified are necessarily the best available for the purpose.] These switches have been shown to work at dilution refrigerator temperatures and commercial products exist for cryogenic use. Unlike the cryogenic solenoid-based switches, these switches are very compact, available as a square surface mount package approximately 2.5 mm on a side and can be integrated on circuit boards to make a wide range of possible systems and products.  To control the RF lines of these switches, 90 volts is required.  For room temperature systems, these signals can be provided by conventional built-in serial communication electronics. Our Arduino-based controller allows experimenters to cheaply and easily control MEMS-based switches, both in a dilution refrigerator and at room temperature.

## 2. Hardware description

**a**.

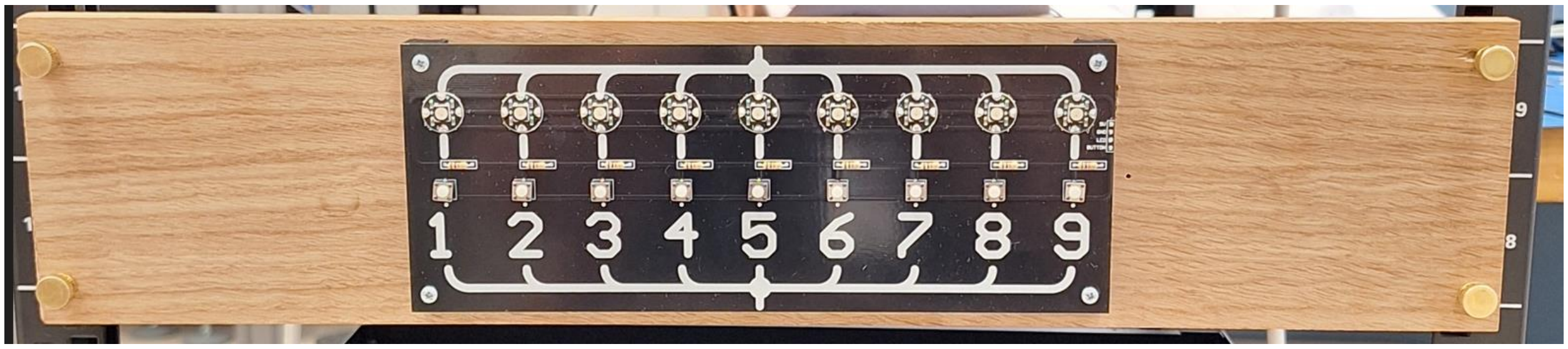

**b.**

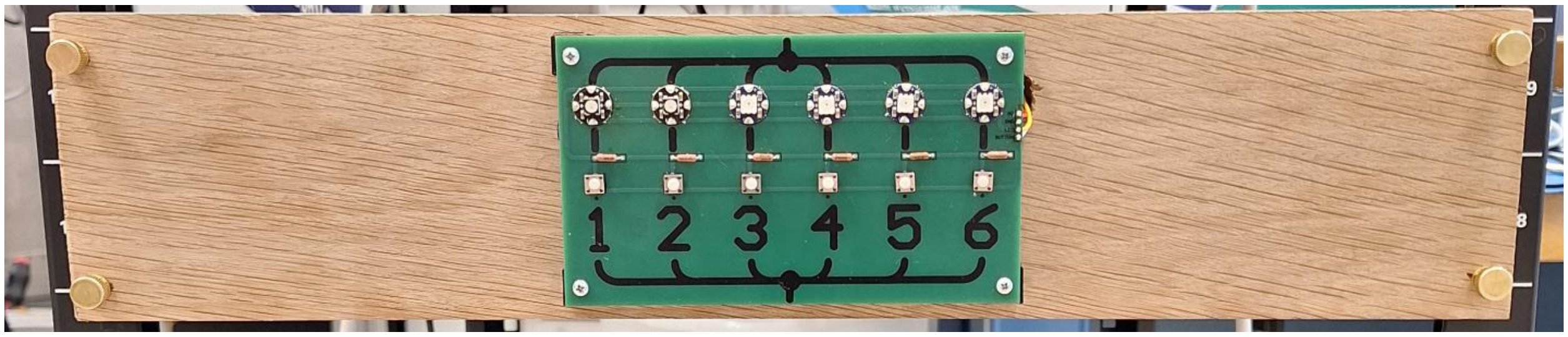

**c.**

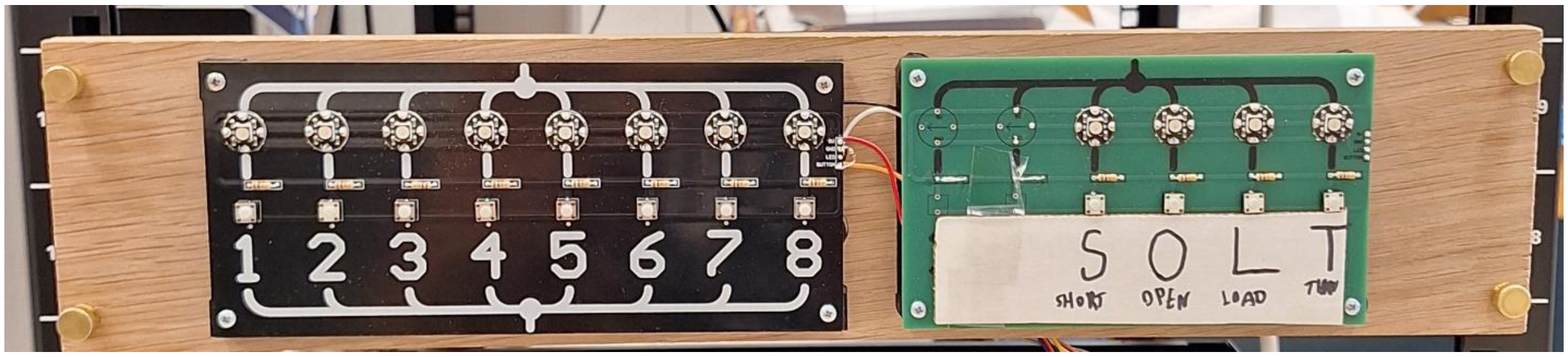

**Figure 2: a.** 9-port switch controller in a 19-inch instrument rack. **b.** SP6T controller for COTS (Commercial Off The Shelf) switches. **c.** SP8T controller with built in electronic calibration (to be documented elsewhere). Four knurled brass screws mount the wood board to the rack.

The switch controller consists of a 19 inch (48.26 cm) rack mount panel just over 2U in height, or 4 inches (100 mm) with a control panel mounted on the front and a metal enclosure and Arduino mounted to the back. The whole apparatus is mounted on a piece of wood cut and drilled to mount in the rack, with appropriate mounting holes for 10-32 rack mount screws. The controller continuously monitors the USB port and the manual push buttons, allowing the user to smoothly transition between automated and manual operation.

The control panel (see Figure 2) consists of a printed circuit with buttons and programmable multi-color LED pixels, and appropriate labeling to make the user interface clear by inspection. The control panel can be redesigned for any switch topology or function, with the labeling modified to indicate the function and lights and switches located to match their position on the switch network. These boards can be redesigned and replaced in less than one week, for under $10 per board for rapid prototyping. We recommend using a black solder mask and white silk screen to achieve maximum color contrast for the panel and using graphics to indicate switch topology in as clear a way as possible for your design, as shown in figure 2.

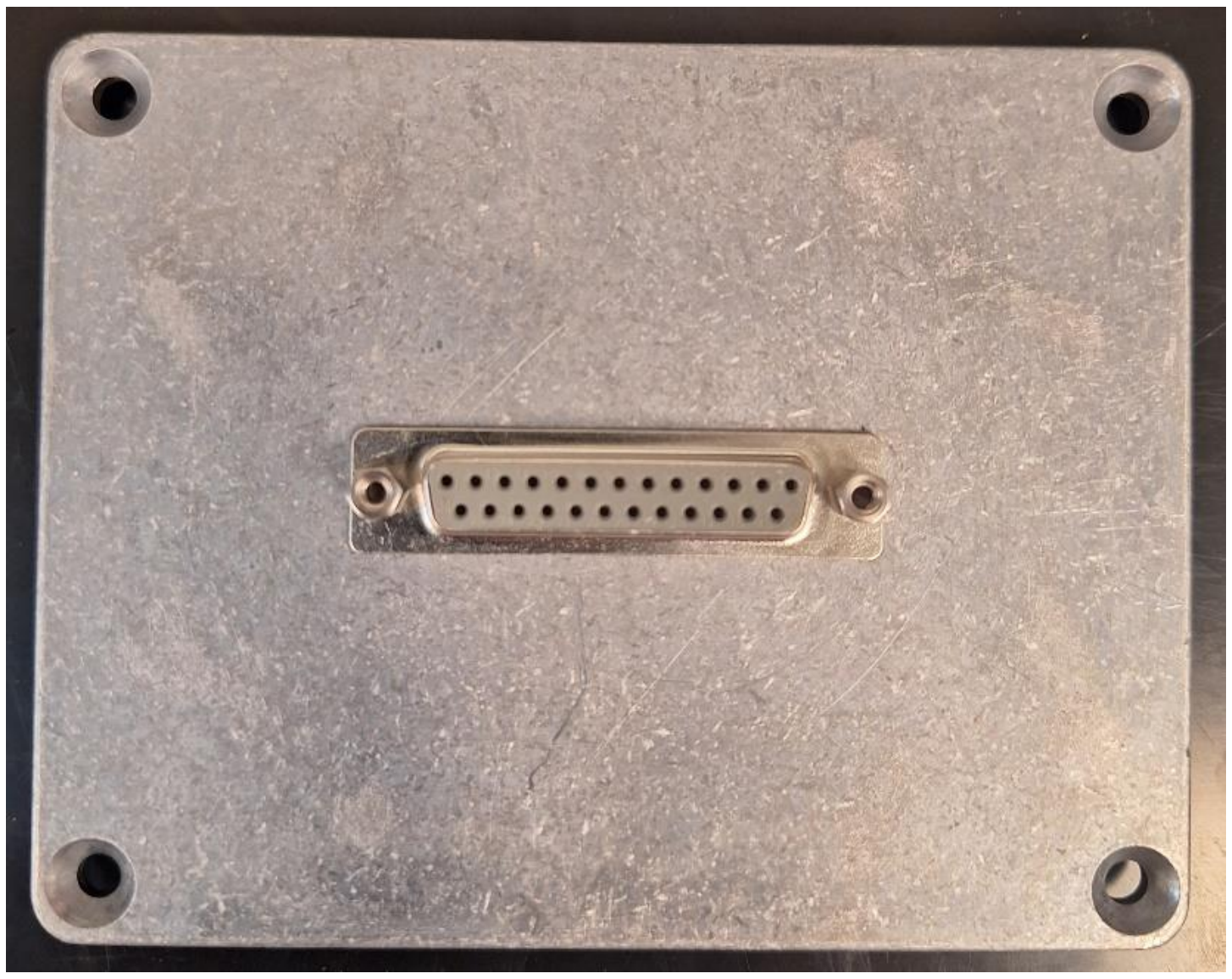

**Figure 3.** Diecast aluminum enclosure with 25 pin DSUB connector and hex standoffs that thread into the lid.

The metal enclosure on the back side of the wood panel isolates the 90 V from the human operator and any other metal objects that might be in the vicinity (see Figure 3). This enclosure can work for any of a variety of options for boosting the voltage from the 5 V signals from the Arduino to the 90 V needed to drive the MEMS switches. A slot for a 25 DSUB connector is machined in the lid of the enclosure, along with 4-40 threaded holes for offset screws which are used to connect a DSUB cable to the box. The main body of the diecast aluminum enclosure has machined in it a square array of 4-40 threaded holes that are used to attach the box to the wood board, as well as threaded holes for mounting the voltage boost circuit board to the inside of the box and a slot milled out of the side for cables to connect from the Arduino board into the board inside the box. The contents of the metal enclosure depend on whether one is using a commercial product that comes with a DC-DC converter up to 90 V or if a custom circuit needs to be fabricated for control of the 90 V. For completeness, we describe a method for building custom controllers with relays and also for building a system using the commercial off-the-shelf (COTS) DC-DC converters which ship with commercial switch systems.

Inside the diecast aluminum enclosure, there is a circuit board called “DB25-relay-HV-control” with a set of 20 relays that can be used to connect a shared power line to any of 20 different pins on the 25-pin DSUB connector. These relays are Comus part number 3570.1331.053 relays which are actuated with a 5 V output from the Arduino (drawing about 10 mA) and can allow for up to 150 V and up to 1 A of current to flow. Note that the part number 3570.1331.053 with a “3” at the end as opposed to a “1” specifies that there is a diode included in the package, which is important to protect the Arduino from driving an inductive load that can create a back electromotive force and damage the Arduino. This board could be re-designed for up to 24 relays and control lines (utilizing all 24 standard DC lines that go into many dilution refrigerators) if needed for some future application; the board could also be expanded to have two DSUB connectors, doubling the number of lines.

The Arduino UNO board (the UNO is a specific model of Arduino board produced by multiple vendors) is mounted to the wood front panel using a three-dimensional (3D) printed bracket called “Arduino-bracket”, which has clearance holes for a pair of #6 wood screws and a set of three 4-40 thread inserts for attaching the Arduino UNO. The same hole pattern and bracket work for mounting the Arduino Mega, which is needed for systems with larger numbers of control lines than 16. An Arduino shield board called “Arduino-to-headers-shield” plugs into the Arduino board via 4 headers, according to the standard header layout of Arduino UNO-like boards. Another set of headers on the top of that board connects to the relay board in the metal enclosure by means of a set of short cables that feed through the slot in the side of the metal box. The shield board also has a 100 kΩ pull-down resistor for the resistor ladder, which connects to the buttons, and a 330 Ω series resistor and 1000 μF capacitor for driving the programmable LEDs.

The control buttons connect a single analog input on the Arduino to different voltage levels, which vary from 5.5 V to ground, by using a ladder of 10 kΩ resistors (see Figure 4). The Arduino code measures the voltage on the analog input twice, with a short delay in between, to verify that a button is indeed being held down and to not respond to transient voltages. This technique of using a resistor ladder to allow many buttons to be read with a single pin can be scaled up or down for as many buttons as needed, up to the uncertainty of the resistor values.

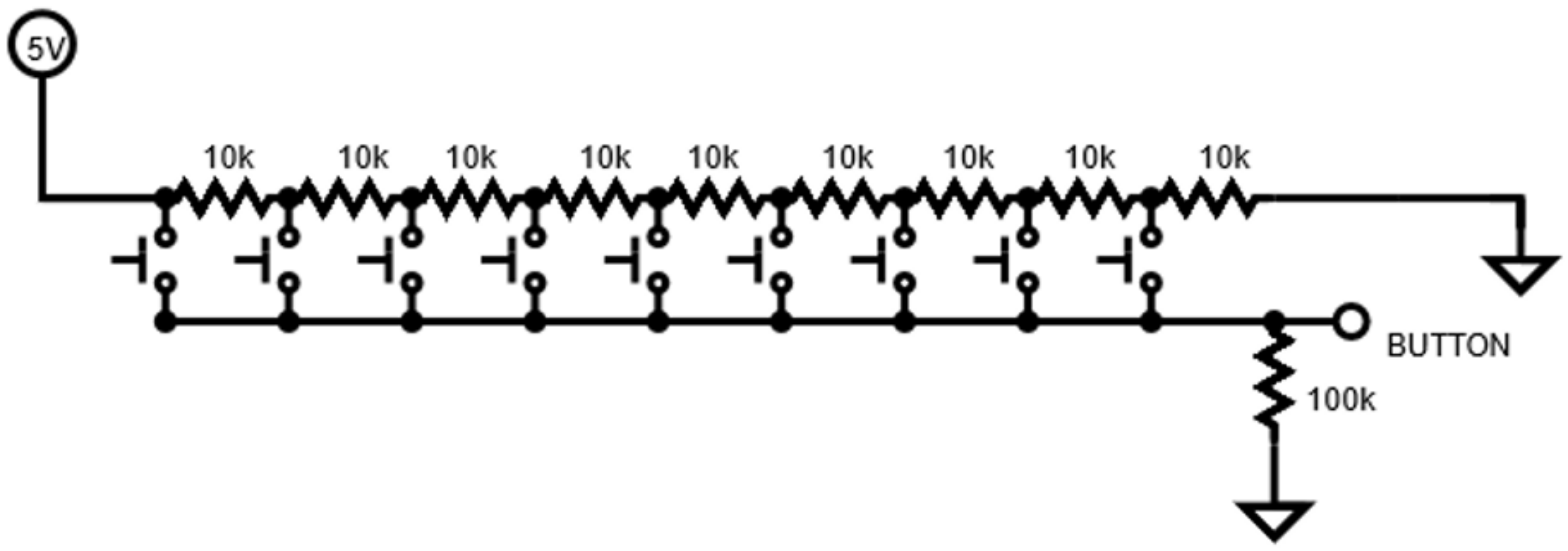

**Figure 4:** A ladder of 10 kΩ resistors with a 100 kΩ pull-down resistor is used to create a many-button control panel with a single analog line. For each new configuration with a different set of buttons, we adjust the threshold values for the analog voltage appropriately for the number of resistors. While this is easy enough to calculate, we have simply measured the value for each button press by streaming the analog value over the serial port and noting values to put in the code. To prevent sensitivity to transient responses, we use nested if statements with a 1 ms time delay to ensure that the voltage is stable in any given button press state.

Like the control buttons, the indicator lights all share a single pin on the Arduino. Each LED is mounted to a small circular circuit board 0.5 inches (12.5 mm) in diameter. These boards are daisy-chained together and can be arranged in whatever scheme is most useful for communicating switch topology to a human operator. The LEDs are programmed in the Arduino code using the NeoPixel open source libraries available from Adafruit). and can be downloaded from inside the Arduino IDE software.

Every part of the MEMSDuino system is modular and can be independently modified for rapid prototyping of a wide range of switch control applications. The system can also accommodate the high current pulses used to control solenoid switches, making the whole apparatus usable for custom-built solenoid-based switching systems as well. MEMSDuino offers numerous advantages, including:

- MEMSDuino is a flexible, inexpensive, and simple platform that can be used to control any of a wide range of existing COTS MEMS RF switches.
- MEMSDuino can be used for rapid prototyping of new designs of MEMS-based RF switch networks.
- In addition to driving MEMS switches, the MEMSDuino can be used for building controllers of systems of solenoid-based switches also used in cryogenic RF test systems.
- MEMSDuino uses a very simple method for building arbitrary control panels of buttons and indicator lights using just one LED drive line and one analog input line.

## 3. Design files summary

| Design file name | File type | Open source license | Location of the file |
|---|---|---|---|
| 6_port_controller.ino | Arduino source code file | Public Domain | *https://github.com/lafefspietz/MEMSduino* |
| 6button-6eopixel.SchDoc | Altium schematic file | Public Domain | *https://github.com/lafefspietz/MEMSduino* |
| 6button-6neopixel-BOM.xlsx | Bill of Materials(Excel file) | Public Domain | *https://github.com/lafefspietz/MEMSduino* |
| 6button-6neopixel-gerbers.zip | ZIP of Gerber Files used to manufacture board | Public Domain | *https://github.com/lafefspietz/MEMSduino* |
| 6button-6neopixel.PcbDoc | Altium layout file | Public Domain | *https://github.com/lafefspietz/MEMSduino* |
| 6button-6neopixels.PrjPcb | Altium project file | Public Domain | *Https://github.com/lafefspietz/MEMSduino* |
| 6button-6neopixels.PrjPcbStructure | Altium project structure file | Public Domain | *Https://github.com/lafefspietz/MEMSduino* |
| 9-port-controller.ino | Arduino source code file | Public Domain | *Https://github.com/lafefspietz/MEMSduino* |
| 9button-9neopixel-BOM.xlsx | Bill of Materials(Excel file) | Public Domain | *Https://github.com/lafefspietz/MEMSduino* |
| 9button-9neopixel-gerbers.zip | ZIP of Gerber Files used to manufacture board | Public Domain | *Https://github.com/lafefspietz/MEMSduino* |
| 9button-9neopixel.PcbDoc | Altium layout file | Public Domain | *Https://github.com/lafefspietz/MEMSduino* |
| 9button-9neopixel.SchDoc | Altium schematic file | Public Domain | *Https://github.com/lafefspietz/MEMSduino* |
| 9button-9neopixels.PrjPcb | Altium project file | Public Domain | *Https://github.com/lafefspietz/MEMSduino* |
| 9button-9neopixels.PrjPcbStructure | Altium project structure file | Public Domain | *Https://github.com/lafefspietz/MEMSduino* |
| AN-1304-A-box-modifications.SLDDRW | SolidWorks drawing file | Public Domain | *Https://github.com/lafefspietz/MEMSduino* |
| AN-1304-A-box-modifications.SLDPRT | SolidWorks part file | Public Domain | *Https://github.com/lafefspietz/MEMSduino* |
| AN-1304-A-box-modifications.STEP | Machining file | Public Domain | *Https://github.com/lafefspietz/MEMSduino* |
| AN-1304-A-box-modifications.pdf | Drawing | Public Domain | *Https://github.com/lafefspietz/MEMSduino* |
| AN-1304-A-lid-modifications.SLDDRW | SolidWorks drawing file | Public Domain | *Https://github.com/lafefspietz/MEMSduino* |
| AN-1304-A-lid-modifications.SLDPRT | SolidWorks part file | Public Domain | *Https://github.com/lafefspietz/MEMSduino* |
| AN-1304-A-lid-modifications.STEP | Machining file | Public Domain | *Https://github.com/lafefspietz/MEMSduino* |
| AN-1304-A-lid-modifications.pdf | Drawing | Public Domain | *Https://github.com/lafefspietz/MEMSduino* |
| DB25-relay-HV-control-BOM.xlsx | Bill of Materials(Excel file) | Public Domain | *Https://github.com/lafefspietz/MEMSduino* |
| DB25-relay-HV-control-gerbers.zip | ZIP of Gerber Files used to manufacture board | Public Domain | *Https://github.com/lafefspietz/MEMSduino* |
| DB25-relay-HV-control.PcbDoc | Altium layout file | Public Domain | *Https://github.com/lafefspietz/MEMSduino* |
| DB25-relay-HV-control.PrjPcb | Altium project file | Public Domain | *Https://github.com/lafefspietz/MEMSduino* |
| DB25-relay-HV-control.PrjPcbStructure | Altium project structure file | Public Domain | *Https://github.com/lafefspietz/MEMSduino* |
| DB25-relay-HV-control.SchDoc | Altium schematic file | Public Domain | *Https://github.com/lafefspietz/MEMSduino* |
| HV-DCDC-bracket.SLDDRW | SolidWorks drawing file | Public Domain | *Https://github.com/lafefspietz/MEMSduino* |
| HV-DCDC-bracket.SLDPRT | SolidWorks part file | Public Domain | *Https://github.com/lafefspietz/MEMSduino* |
| HV-DCDC-bracket.STL | 3d print file | Public Domain | *Https://github.com/lafefspietz/MEMSduino* |
| HV-DCDC-bracket.pdf | Drawing | Public Domain | *Https://github.com/lafefspietz/MEMSduino* |
| SP6T-COTS-BOM.xlsx | Bill of Materials(Excel file) | Public Domain | *Https://github.com/lafefspietz/MEMSduino* |
| SP9T-Custom-BOM.xlsx | Bill of Materials(Excel file) | Public Domain | *Https://github.com/lafefspietz/MEMSduino* |
| arduino-bracket.SLDDRW | SolidWorks drawing file | Public Domain | *Https://github.com/lafefspietz/MEMSduino* |
| arduino-bracket.SLDPRT | SolidWorks part file | Public Domain | *Https://github.com/lafefspietz/MEMSduino* |
| arduino-bracket.STL | 3d print file | Public Domain | *Https://github.com/lafefspietz/MEMSduino* |
| arduino-bracket.pdf | Drawing | Public Domain | *Https://github.com/lafefspietz/MEMSduino* |
| arduino-to-headers-shield-BOM.xlsx | Bill of Materials(Excel file) | Public Domain | *Https://github.com/lafefspietz/MEMSduino* |
| arduino-to-headers-shield-gerbers.zip | ZIP of Gerber Files used to manufacture board | Public Domain | *Https://github.com/lafefspietz/MEMSduino* |
| arduino-to-headers-shield.PcbDoc | Altium layout file | Public Domain | *Https://github.com/lafefspietz/MEMSduino* |
| arduino-to-headers-shield.PrjPcb | Altium project file | Public Domain | *Https://github.com/lafefspietz/MEMSduino* |
| arduino-to-headers-shield.PrjPcbStructure | Altium project structure file | Public Domain | *Https://github.com/lafefspietz/MEMSduino* |
| arduino-to-headers-shield.SchDoc | Altium schematic file | Public Domain | *Https://github.com/lafefspietz/MEMSduino* |
| cryoelec-bracket.SLDDRW | SolidWorks drawing file | Public Domain | *Https://github.com/lafefspietz/MEMSduino* |
| cryoelec-bracket.SLDPRT | SolidWorks part file | Public Domain | *Https://github.com/lafefspietz/MEMSduino* |
| cryoelec-bracket.STL | 3d print file | Public Domain | *Https://github.com/lafefspietz/MEMSduino* |
| cryoelec-bracket.pdf | Drawing | Public Domain | *Https://github.com/lafefspietz/MEMSduino* |
| front-panel-center-bracket.SLDDRW | SolidWorks drawing file | Public Domain | *Https://github.com/lafefspietz/MEMSduino* |
| front-panel-center-bracket.SLDPRT | SolidWorks part file | Public Domain | *Https://github.com/lafefspietz/MEMSduino* |
| front-panel-center-bracket.STL | 3d print file | Public Domain | *Https://github.com/lafefspietz/MEMSduino* |
| front-panel-center-bracket.pdf | Drawing | Public Domain | *Https://github.com/lafefspietz/MEMSduino* |
| front-panel-edge-bracket.SLDDRW | SolidWorks drawing file | Public Domain | *Https://github.com/lafefspietz/MEMSduino* |
| front-panel-edge-bracket.SLDPRT | SolidWorks part file | Public Domain | *Https://github.com/lafefspietz/MEMSduino* |
| front-panel-edge-bracket.STL | 3d print file | Public Domain | *Https://github.com/lafefspietz/MEMSduino* |
| front-panel-edge-bracket.pdf | Drawing | Public Domain | *Https://github.com/lafefspietz/MEMSduino* |
| rack-mount-board-SP6T.SLDDRW | SolidWorks drawing file | Public Domain | *Https://github.com/lafefspietz/MEMSduino* |
| rack-mount-board-SP6T.SLDPRT | SolidWorks part file | Public Domain | *Https://github.com/lafefspietz/MEMSduino* |
| rack-mount-board-SP6T.pdf | Drawing | Public Domain | *Https://github.com/lafefspietz/MEMSduino* |
| rack-mount-board-SP9T.SLDDRW | SolidWorks drawing file | Public Domain | *Https://github.com/lafefspietz/MEMSduino* |
| rack-mount-board-SP9T.SLDPRT | SolidWorks part file | Public Domain | *Https://github.com/lafefspietz/MEMSduino* |
| rack-mount-board-SP9T.pdf | Drawing | Public Domain | *Https://github.com/lafefspietz/MEMSduino* |
| switch_control_matlab.m | Matlab source code file | Public Domain | *Https://github.com/lafefspietz/MEMSduino* |
| switch_control_python.py | Python source code file | Public Domain | *Https://github.com/lafefspietz/MEMSduino* |

## 4. Bill of materials summary

**SP6T COTS Build BOM**

| descriptior | link to buy | unit price | number | extended ¦ | vendor | vendor pai | manfactur | manufacturer part nu |
|---|---|---|---|---|---|---|---|---|
| 1/2" White | https://oc( | $ 5.40 | 1 | $ 5.40 | Ocooch Hardwoods | | Ocooch Hardwoods | |
| Brass Flare | https://wv | $ 1.62 | 4 | $ 6.48 | McMaster | 92421A645 | | |
| #6 wood s | https://wv | $ 0.07 | 8 | $ 0.56 | Amazon | | | |
| cryoelec-b | https://github.com/la | | 1 | | in house prototype | | | |
| arduino-br | https://github.com/la | | 1 | | in house prototype | | | |
| front-pane | https://github.com/la | | 2 | | in house prototype | | | |
| Brass threa | https://wv | $ 0.08 | 7 | $ 0.57 | 3dJake | RUT-GE-4- | ruthex | GE-4-40x57-001 |
| 0-80 Brass | https://wv | $ 0.47 | 4 | $ 1.88 | Amazon | B08QJJDC\ | E-Z LOK | 240-000-BR |
| Zinc-Platec | https://wv | $ 0.02 | 9 | $ 0.16 | McMaster | 90272A106 | | |
| Diecast alu | https://wv | $ 14.90 | 1 | $ 14.90 | Digikey | 377-2395- | Bud Indust | AN-1304-A |
| Male-Fema | https://wv | $ 3.73 | 2 | $ 7.46 | McMaster | 91075A462 | | |
| Brass Sock | https://wv | $ 1.20 | 2 | $ 2.40 | McMaster | 93465A113 | | |
| Brass Was | https://wv | $ 0.05 | 2 | $ 0.10 | McMaster | 92916A320 | | |
| 6button-6r | https://git| | $ 5.98 | 1 | $ 5.98 | PCBWay | | | |
| arduino-to | https://git| | $ 1.82 | 1 | $ 1.82 | PCBWay | | | |
| 100k axial | https://wv | $ 0.10 | 1 | $ 0.10 | Digikey | RNF14FTD | Stackpole | RNF14FTD100K |
| 10k axial r | https://wv | $ 0.05 | 9 | $ 0.45 | Digikey | 10.0KXBK- | YAGEO | MFR-25FBF52-10K |
| 330 ohm a | https://wv | $ 0.10 | 2 | $ 0.20 | Digikey | CF18JT330 | Stackpole | CF18JT330R |
| 1000 uF el | https://wv | $ 0.51 | 1 | $ 0.51 | Digikey | 1189-1719 | Rubycon | 16PK1000MEFC10X12 |
| Momentar | https://wv | $ 0.29 | 6 | $ 1.74 | Digikey | EG2532TB | E Switch | TL59NF160Q |
| JUMPER W | https://wv | $ 2.10 | 1 | $ 2.10 | Digikey | 1568-1512 | Sparkfun | PRT-12795 |
| DB25 pane | https://wv | $ 2.16 | 1 | $ 2.16 | Digikey | 225ME-NE | NorComp | 171-025-103L001 |
| Arduino UI | https://wv | $ 16.90 | 1 | $ 16.90 | Digikey | 1738-1228 | DFRobot | DFR0216 |
| USB A to U | https://wv | $ 1.64 | 1 | $ 1.64 | Digikey | AE1493-NI | Assmann V | AK672/2-2 |
| Neopixel p | https://wv | $ 1.25 | 6 | $ 7.50 | Adafruit | 1559 | Adafruit | 1559 |
| 2 wire hea | https://wv | $ 0.60 | 1 | $ 0.60 | Amazon | | | |
| 4 wire hea | https://wv | $ 0.79 | 9 | $ 7.11 | Amazon | | | |
| 8 wire hea | https://wv | $ 0.99 | 2 | $ 1.98 | Amazon | | | |
| 0.1" heade | https://wv | $ 0.09 | 3 | $ 0.27 | Amazon | | | |
| | | | | | | | | |

**SP9T Custom Build BOM**

| descriptio | link | unit price | number | extended | vendor | vendor pa | manfactur | manufacturer part nu |
|---|---|---|---|---|---|---|---|---|
| 1/2" Whit | https://oc | $ 5.40 | 1 | $ 5.40 | Ocooch Hardwoods | | Ocooch Hardwoods | |
| Brass Flar | https://ww | $ 1.62 | 4 | $ 6.48 | McMaster | 92421A645 | | |
| #6 wood s | https://ww | $ 0.07 | 10 | $ 0.70 | Amazon | | | |
| HV-DCDC- | https://github.com/l | | 1 | | in house prototype | | | |
| arduino-b | https://github.com/l | | 1 | | in house prototype | | | |
| front-pane | https://github.com/l | | 2 | | in house prototype | | | |
| front-pane | https://github.com/l | | 2 | | in house prototype | | | |
| Brass thre | https://ww | $ 0.08 | 11 | $ 0.89 | 3dJake | RUT-GE-4- | ruthex | GE-4-40x57-001 |
| Zinc-Plate | https://ww | $ 0.02 | 11 | $ 0.20 | McMaster | 90272A106 | | |
| Diecast alu | https://ww | $ 14.90 | 1 | $ 14.90 | Digikey | 377-2395- | Bud Indus | AN-1304-A |
| Male-Fem | https://ww | $ 3.73 | 2 | $ 7.46 | McMaster | 91075A462 | | |
| Brass Sock | https://ww | $ 1.20 | 2 | $ 2.40 | McMaster | 93465A113 | | |
| Brass Was | https://ww | $ 0.05 | 2 | $ 0.10 | McMaster | 92916A320 | | |
| 5V-12V bo | https://ww | $ 11.99 | 1 | $ 11.99 | Amazon | B01ID90K4 | KUNCAN | KC-SU-10-2 |
| 12V to HV | https://ww | $ 11.39 | 1 | $ 11.39 | Amazon | B09D93QN | QINIZX | NIXIE-HV380V |
| heat shrin | https://ww | $ 1.89 | 1 | $ 1.89 | Digikey | 19-SFTW2 | 3M | SFTW203 1/16" BL SP |
| 9button-9 | https://git | $ 5.98 | 1 | $ 5.98 | PCBWay | | | |
| arduino-to | https://git | $ 1.82 | 1 | $ 1.82 | PCBWay | | | |
| DB25-relay | https://git | $ 1.82 | 1 | $ 1.82 | PCBWay | | | |
| 100k axial | https://ww | $ 0.10 | 1 | $ 0.10 | Digikey | RNF14FTD | Stackpole | RNF14FTD100K |
| 10k axial r | https://ww | $ 0.05 | 9 | $ 0.45 | Digikey | 10.0KXBK- | YAGEO | MFR-25FBF52-10K |
| 330 ohm a | https://ww | $ 0.10 | 2 | $ 0.20 | Digikey | CF18JT330 | Stackpole | CF18JT330R |
| 1000 uF el | https://ww | $ 0.51 | 1 | $ 0.51 | Digikey | 1189-1719 | Rubycon | 16PK1000MEFC10X12 |
| Momenta | https://ww | $ 0.29 | 9 | $ 2.61 | Digikey | EG2532TB | E Switch | TL59NF160Q |
| 1 Mohm 1 | https://ww | $ 0.02 | 20 | $ 0.40 | Digikey | 311-1.00M | YAGEO | RC1206FR-071ML |
| DB25 boar | https://ww | $ 1.17 | 1 | $ 1.17 | Digikey | AE10940-N | Assmann W | A-DF 25 PP/Z |
| RELAY REE | https://ww | $ 1.42 | 20 | $ 28.40 | Digikey | 1835-1117 | Comus Int | 3570-1331-053 |
| Arduino U | https://ww | $ 16.90 | 1 | $ 16.90 | Digikey | 1738-1228 | DFRobot | DFR0216 |
| USB A to U | https://ww | $ 1.64 | 1 | $ 1.64 | Digikey | AE1493-NI | Assmann W | AK672/2-2 |
| Neopixel p | https://ww | $ 1.25 | 9 | $ 11.25 | Adafruit | 1559 | Adafruit | 1559 |
| 2 wire hea | https://ww | $ 0.60 | 1 | $ 0.60 | Amazon | | | |
| 4 wire hea | https://ww | $ 0.79 | 9 | $ 7.11 | Amazon | | | |
| 8 wire hea | https://ww | $ 0.99 | 2 | $ 1.98 | Amazon | | | |
| 0.1" heade | https://ww | $ 0.09 | 3 | $ 0.27 | Amazon | | | |

## 5. Build instructions

The build begins by constructing the wood panel. Hardwood (White Oak) boards are purchased in 24" x 4" x 0.5" pieces, which are cut down to 19 inches. This cut should be precise enough to fit in a 19-inch rack and have room for the 10-32 mounting holes. Print out the templates from the .pdf of the layout of the wood panel, cut them out, then tape the two halves together to make one unified template that is exactly the size of the 19x4 inch board. Use this template to punch dimples into the wood where holes will be drilled, then drill small pilot holes and then larger holes, using the dimensions specified in the template (see Figure 5). Drill out a slot as shown on the template; this can be done by drilling a series of ¼ in holes. We recommend using sandpaper and/or a file to remove splinters from all edges.

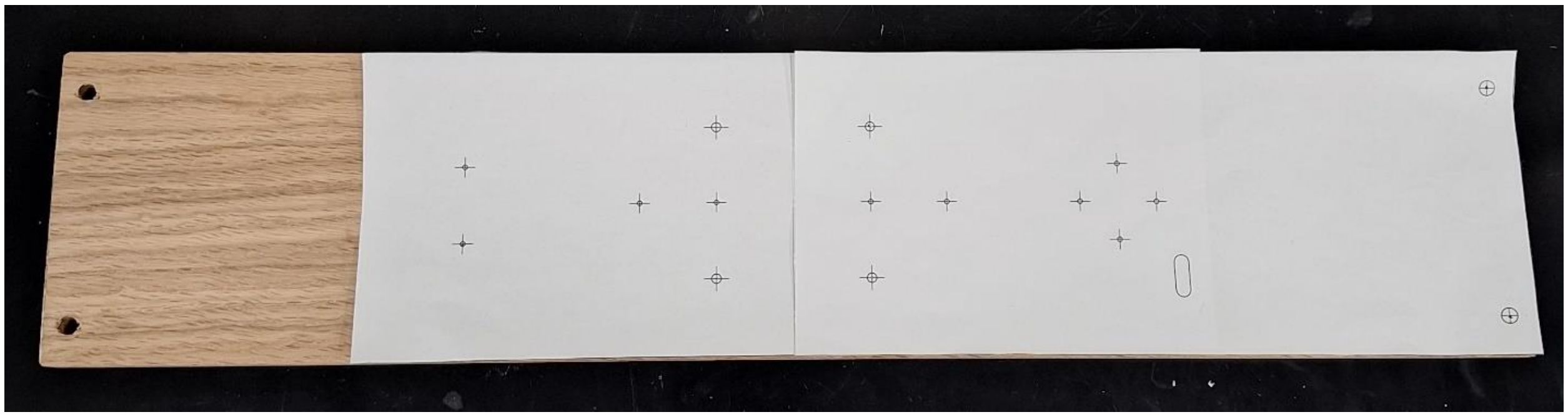

**Figure 5.** The panel hole layouts for both the SP9T and SP6T controllers are traced with a sharp tool using printouts from the drawings printed in 1 to 1 scale, cut out, and taped together over the wood. This method of transferring hole positions and cuts to wood requires only simple hand tools, tape, and scissors and a sharp implement.

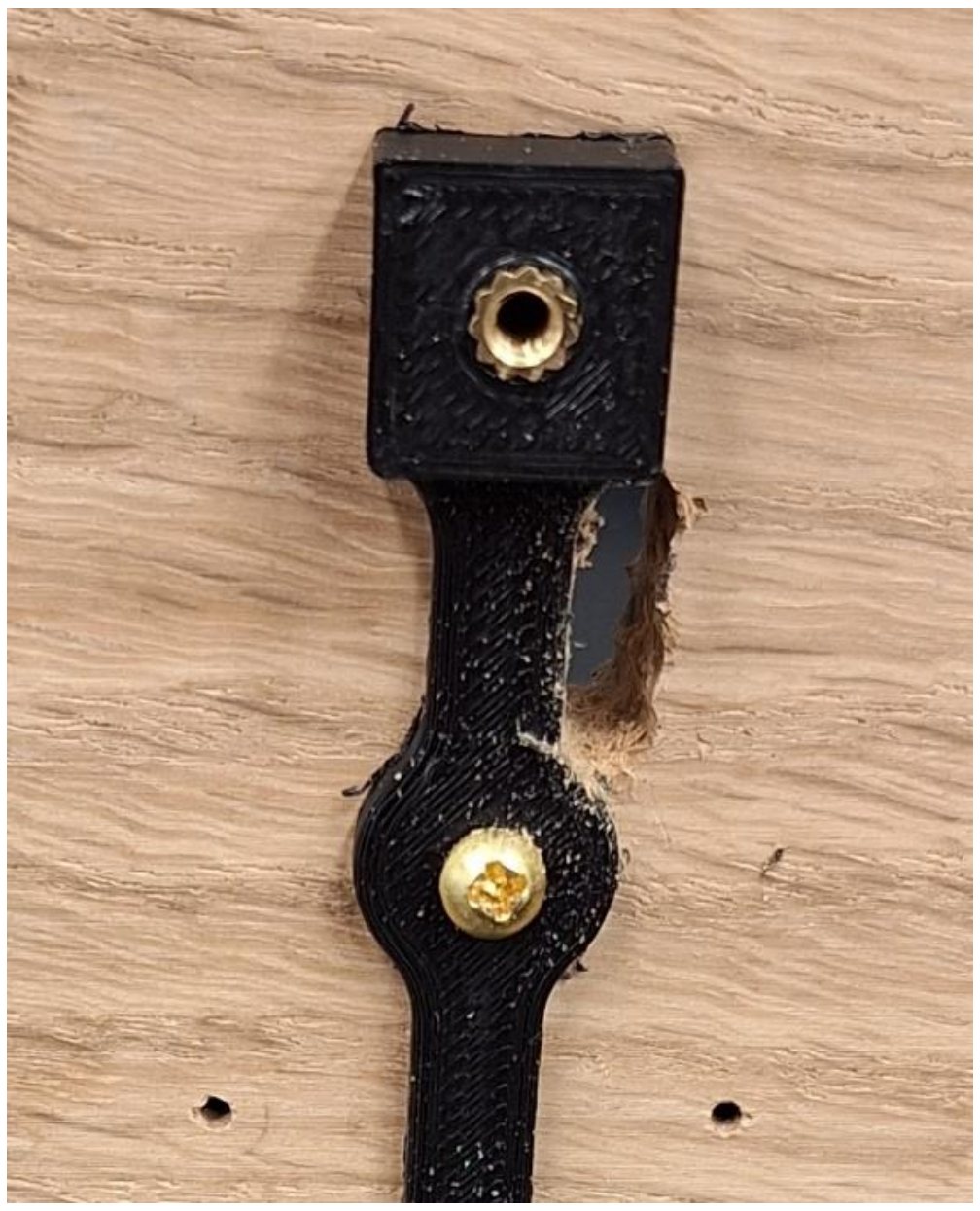

**Figure 6.** Cutout in the wood board for the connector on the 4-wire cable, which connects the control panel with the Arduino Shield board. Although a milling machine could be used, a few holes drilled with a hand drill and connected manually work well enough.

Purchase the Bud Industries AN-1304-A diecast aluminum enclosure and either order the modifications or make them yourself if you have access to a machine shop with a milling machine. To get it machined by an outside shop, send them the .STEP file and the .pdf included in the repository for this paper. The original SolidWorks drawings are also included in the repository so that the design can be modified as needed. Note that the 3D files include only the pre-tapped hole sizes for the tapped holes.

Place the DB25 board mount connector (Assmann WSW Components A-DF 25 PP/Z) from the outside of the D-cutout in the box lid and screw the threaded 4-40 hex offsets into the holes on the edges of the connector. This *must* be done *before* soldering the relay board to the connector! If this is soldered to the board first, it will no longer be possible to mount the connector in the D-cutout in the lid.

To fully assemble all the 3D printed components, print out one each of *Arduino-bracket.STL*, and *HV-DCDC-bracket.STL*, and two each of *front-panel-center-bracket.STL* and *front-panel-edge-bracket.STL*. Note that for the SP6T switch configuration, one can print out just the edge panel brackets (see Figure 6) and skip the center brackets; this is somewhat a matter of preference, to decide how much mechanical support you want the control panel circuit board to have. All these 3D printed components, other than the optional center brackets, require the insertion of 4-40 brass thread inserts. The insertion process is accomplished by setting a soldering iron to about 250 C and manually pressing vertically on the insert although it is easier if you have a vertical press that can also hold the soldering iron in place. A pair of pliers can be useful to remove the soldering iron if it gets stuck in the insert. Thread inserts can be purchased from both Amazon and from 3D printer suppliers and they are critical for making 3D printed brackets with threads for small screws.

With the 3D printed parts printed and the brass thread inserts added, we can screw the Arduino bracket, the inner and outer front panel mounting brackets, and the diecast aluminum box to the wood board. The mounting bracket for the DC-DC converter is also then screwed into place inside the aluminum box.

The room temperature controller has three circuit boards, all of which can be fabricated from the most cost-effective circuit boards. We note that it is possible to get any color for both solder mask and silk screen now, and that using a black solder mask with white silk screen gives the maximum color contrast for the control panel board. To order circuit boards without changing the files, simply use the *.ZIP* files which contain the Gerber files used by printed circuit board manufacturers to make boards. Be sure to specify the color of both the solder mask (black) and the silk screen (white), and to select “lead free”, and to set the sizes of the boards when generating the quote. The sizes of the boards are specified in the list of board files and bill of materials.

To build the control panel circuit board, we remove the pins from header strips using a pair of needle nose pliers, then place those into the holes between the NeoPixel boards and the control panel board, being careful to orient the arrow on the two circuit boards which indicates signal direction and ensure plus and minus power leads connect correctly. When these are gently placed in the right position, carefully solder them into place, taking care not to disturb the somewhat delicate surface mount WS2812 LEDs on the NeoPixel daughter boards. Add all the 10 kΩ resistors and buttons to the board and solder them. Finally, to connect that board to the rest of the system cut off one end of a 4-wire header cable and solder the 4 wires into the row of 4 plated holes at the side of the board. Alternatively, this can have a header soldered to it and the cable can be plugged in and out of that connector. This can be more mechanically robust but adds a protruding object to the front panel which may be undesirable.

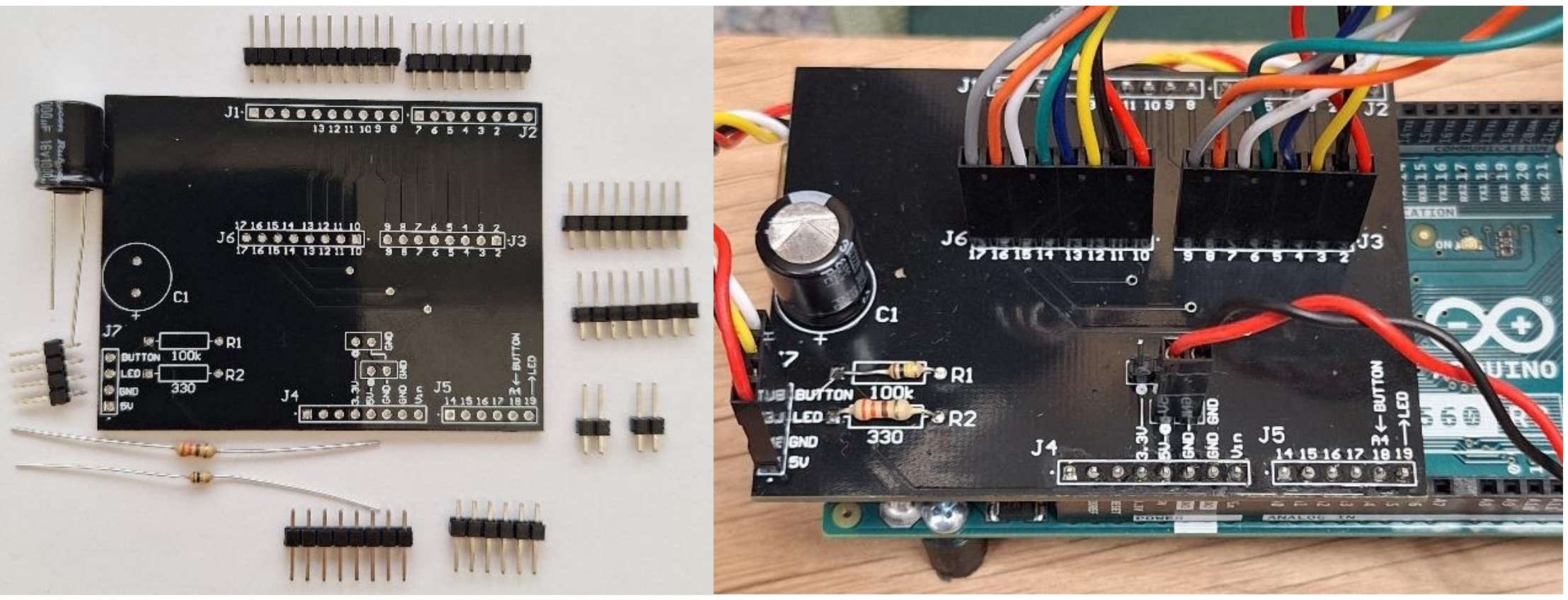

**Figure 7.** Arduino shield board assembly, showing the components before assembly (left) and the full assembly with the Arduino UNO board and all cables connected (right).

To assemble the Arduino shield board, first break off the headers from the 0.1 in (0.254 cm) header strips with a pliers into 1 section with 6 pins, one with 10 pins, and two with 8 pins (see Figure 7). Insert those headers into the header socket strips on the Arduino UNO, then place the board face up on the header strips, and solder all of them into place, being careful to check that no pins have been missed. It is easy to miss a pin here and end up with a bad connection later. Add the headers that are on the top of the board as shown in figure 7, and solder in the two resistors and the capacitor. Be careful to note the polarity on the capacitor; if mounted backwards, it will not work and can catch fire or explode.

To assemble the relay board (see Figure 8), be sure to solder it to the captive 25-pin DSUB connector last after the rest of the board has been assembled. Start the assembly by soldering the surface mount resistors to the top of the board by putting a solder blob on one side first, then, holding the resistor with tweezers in one hand and the soldering iron in the other, melt the blob and place the resistor. Then with one end anchored, you can use the soldering iron and a piece of solder to solder the other end of the resistor. Be sure to check all resistors to make sure you did not miss one of the second solder joints when you're done. These resistors are needed to pull the outputs of the lines down when they're not connected to the high voltage by a relay. When the resistors are mounted, solder in the relays on the back side, being careful to note which end is pin 1 on the relay and on the board. Solder in the headers also on the back side, the same side as the relays. With all these parts together, finally, you can solder the board into a captivated position on the lid of the aluminum enclosure.

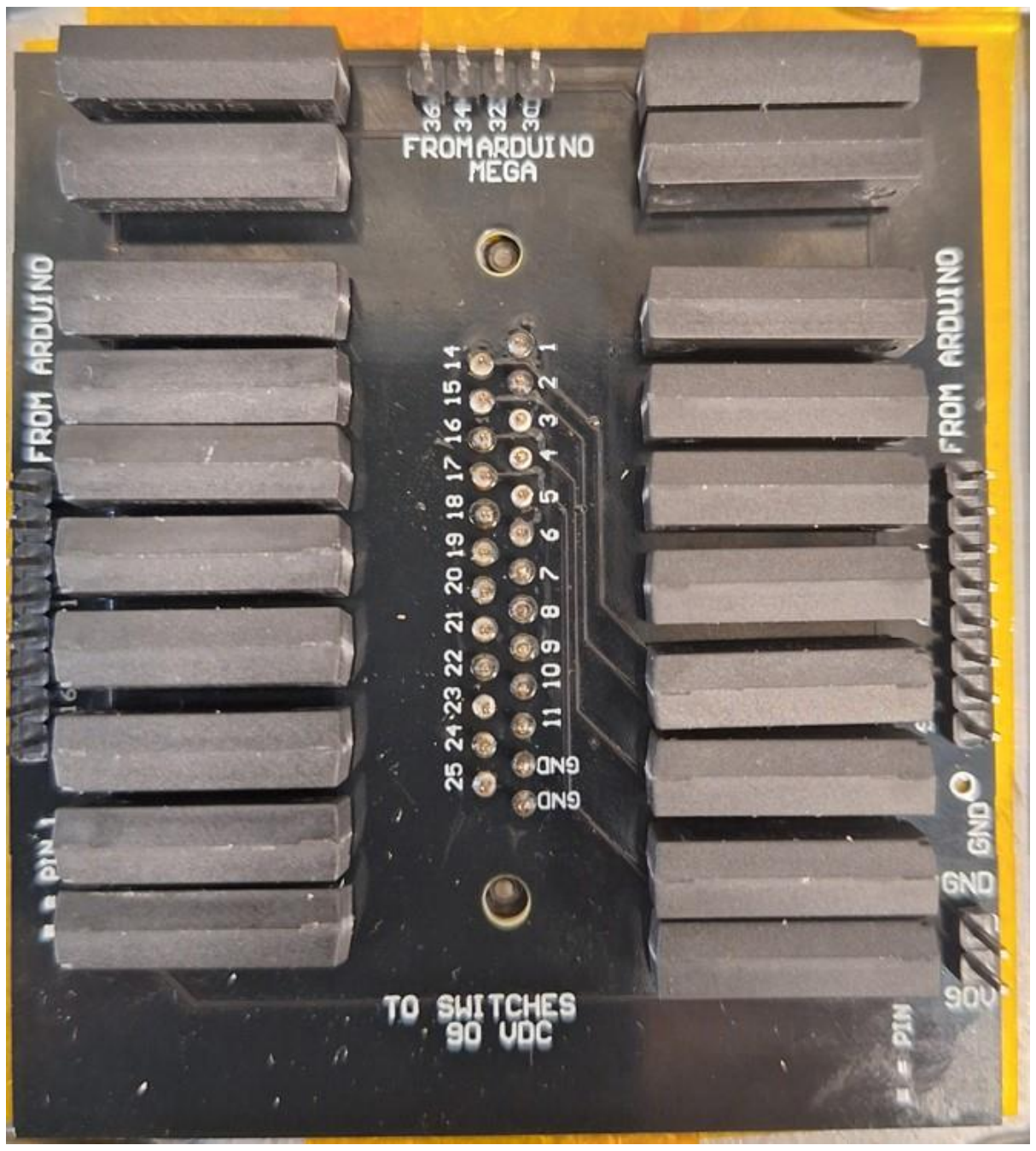

**Figure 8.** Circuit board DB25-relay-HV-control connecting the DSUB connector on the lid of the aluminum enclosure to a set of relays all connected to the 90 V line. The solder joint between the board and the DSUB must be made last, after everything else has been assembled. Once completed, the board is captive on the metal box lid.

Note that this is only for the custom-built switch matrices. For COTS switches, which are shipped with step-up power supplies (see Figure 9), a 25-pin DSUB connector with solder cups should be used to which jumper wires are soldered. In this case, the step-up board is placed inside the aluminum enclosure with double-sided sticky tape, and jumpers are screwed into the screw terminals and connected to the Arduino output on one side and the DSUB connector on the other side.

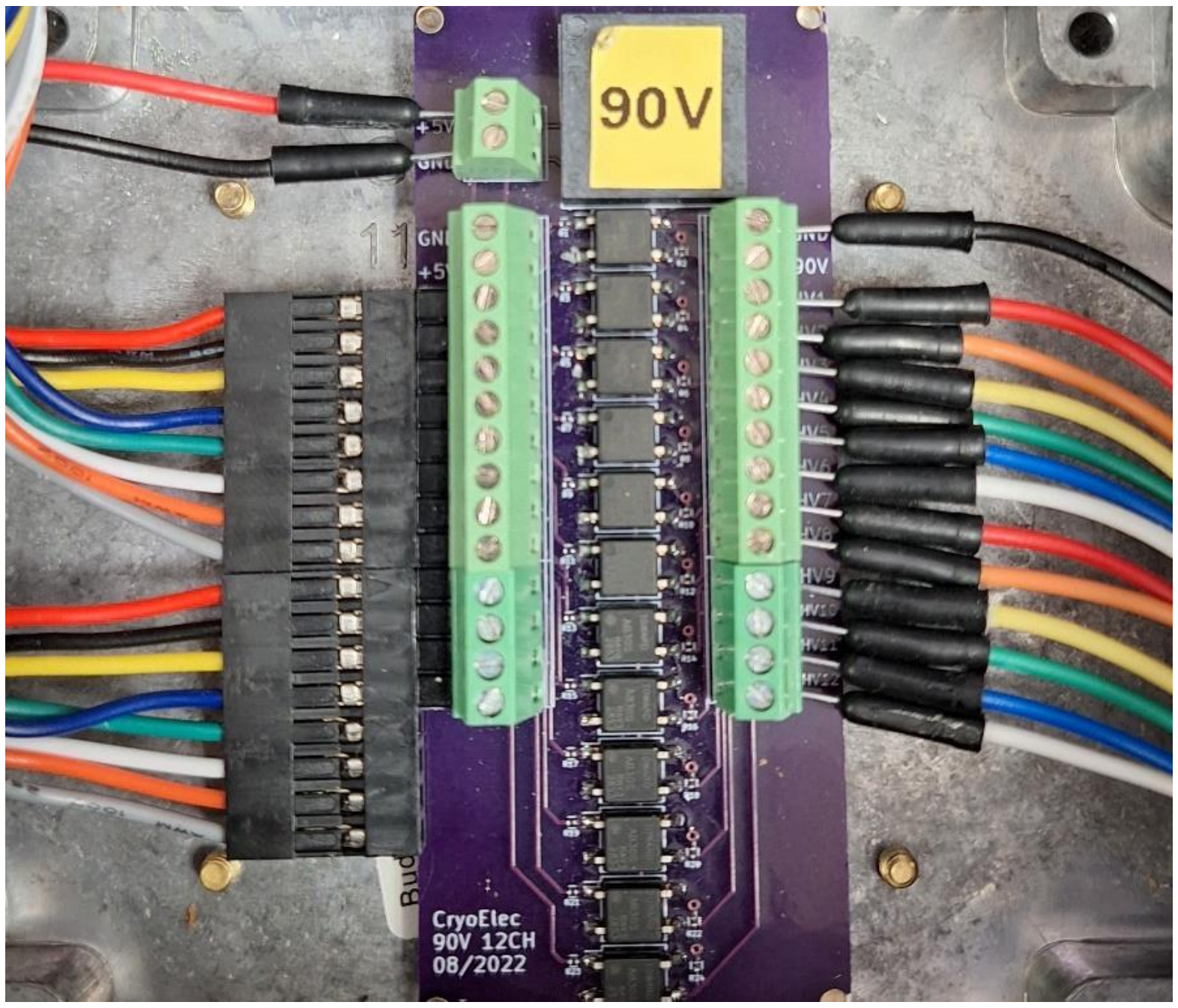

**Figure 9.** COTS 5 V to 90 V boost board in diecast aluminum enclosure. A combination of COTS jumpers can be used to connect this board with the DSUB on the output side and to the Arduino UNO shield board on the other side.

For the custom-built switch systems, we must also supply the 90 V power supply line which connects to all the relays. This can be done with two step-up converters, one from 5 V to 12 V and one from 12 V to a variable voltage up to 350 V. We assemble this system by purchasing the two DC-DC step-up converters and connecting them together in series with a set of 2-wire header cables with solder joints and heat shrink tubing. Carefully check the output voltage with a voltmeter with range over 90 V as you turn the adjustment potentiometer with a small flat head screwdriver until you get close to 90 V. Carefully verify that black is minus and red is plus and connect the power between the Arduino shield and the relay boards. Connect the 8-wire header cables, carefully making sure the orientation matches the pin numbers on the silk screens of the two boards.

With all the physical parts assembled, connect a USB A to USB B cable to a computer running the Arduino IDE. Be sure that the Adafruit NeoPixel library is installed, searching for the library and installing it using the menus in the IDE. Program and test the system. Push all the buttons and check that the switches switch. Next, copy and paste the code snippets for either Python or Matlab to control over the serial port in whatever language you use for taking data in the rest of your quantum radio frequency systems.

## 6. Operation instructions

To use the MEMSDuino controller, connect the controller to the dilution refrigerator (cryostat) using a cable that converts from the DSUB on the MEMSDuino box to the appropriate connector on the cryostat. Connect the MEMS switch matrix to the cryostat by means of a breakout board or adapter cable, typically to a 25-pin micro-D connector.

Use a USB A to USB B cable to connect the MEMSDuino to the same computer used for taking data. Once it has power, push buttons and observe lights to make sure that the manual control works. To test the software control, open the Arduino IDE, make sure the Arduino is visible, and take note of which COM port it is on. This can sometimes involve unplugging and re-plugging in the Arduino to make sure it is clear which port it is on and that it is being recognized. Once the Arduino has been recognized by the IDE and selected in the menu (see Figure 10), you can open the serial monitor in the Arduino IDE and try sending commands to the device. The commands are numbers corresponding to the port selected, so 1 through 9 for the SP9T and 1 through 6 for the SP6T. Be sure to set the serial monitor to 9600 baud to match the baud rate of the Arduino code (this can be changed if you wish, but you must change it in both the Arduino code and also in the control code.)

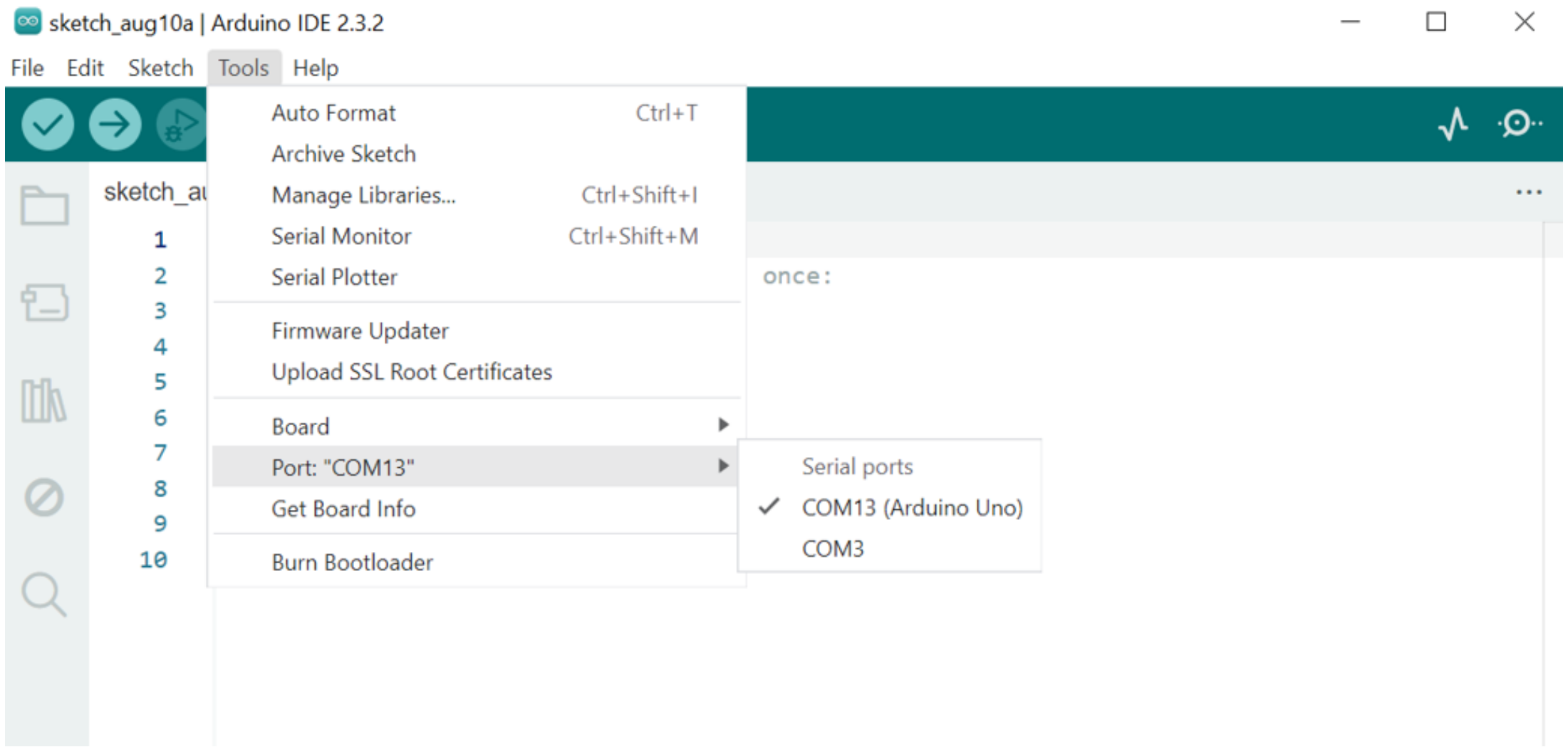

**Figure 10.** Finding the COM port for the Arduino circuit board connected over USB using the Arduino IDE

To control the switches with Python, use the following code

```
import serial
import serial.tools.list_ports
ports = serial.tools.list_ports.comports()

for port, desc, hwid in sorted(ports):
        print("{}: {} [{}]".format(port, desc, hwid))
```

to get a list of com ports available.

When the COM port for the Arduino has been determined, open the port with the following command:

```
SP9T_serial = serial.Serial('COM18', 9600)  # open serial port
```

Then to select, for example, port 2 on the switch, use the following command in Python:

```
SP9T_serial.write(bytes(str(2), encoding='utf-8'))
```

Replace the number 2 with desired port number to be switched. With that code snippet, it is possible to then integrate switch control into the rest of your radio frequency quantum experiments.

## 7. Validation and characterization

Our group at NIST has fabricated several of these instruments to control both custom-built switches we have designed as well as two different COTS switch packages. We are developing more fully integrated systems for cryogenic calibration and characterization of devices, measuring all 4 parameters in the scattering matrix for two terminal devices in a dilution refrigerator. For prototyping a new configuration, an existing front panel control board is used, and paper, pen and tape are used to change the labels of the buttons and lights on-the-fly. We have found that the ability to immediately push buttons and see lights for controlling new switch configurations aids in the development process. In summary:

- We have demonstrated the MEMSDuino controller with both SP6T and SP3T COTS commercial switches.
- We have demonstrated the MEMSDuino controller with both our custom-built SP8T switches and our dual SP9T switches we built for high throughput measurement of quantum devices and cryogenic calibrations

## CRediT author statement

**Lafe Spietz**: Writing - Original Draft , Data Curation, Validation, Writing - Review & Editing, Software, Conceptualization, Investigation **Adam Sirois**: Writing - Review & Editing ,Project administration, Supervision, **Nathan Flowers-Jacobs**: Conceptualization, Validation, **Steve Waltmann**: Conceptualization, Validation,, **Peter Hopkins**: Writing - Review & Editing ,Funding acquisition, Project administration, Supervision **Sam Benz**: Funding acquisition, Project administration, Supervision, Writing - Review & Editing

## Acknowledgments

Funding for this work was provided in part by QED-C